\documentclass[%
 aps,
 prb,%
 amsmath,amssymb,
 reprint,%
superscriptaddress,%
twocolumn,
]{revtex4-2}

\usepackage{physics}
\usepackage{graphicx}
\usepackage{dcolumn}
\usepackage{bm}
\usepackage{dsfont}
\usepackage[dvipsnames]{xcolor}

\begin{document}
\title{Excitation of quasiparticle pairs in superconducting nanodevices by 1/f noise}

\author{Clare C.\ Yu}
\affiliation{Department of Physics and Astronomy, University of California, Irvine, Irvine, CA 92697, USA}

\author{Marcus\ C.\ Goffage}
\affiliation{School of Physics, The University of New South Wales, Sydney, NSW 2052, Australia}

\author{Yifan Wang}
\affiliation{School of Physics, The University of New South Wales, Sydney, NSW 2052, Australia}

\author{Abhijeet\ Alase}
\affiliation{{Department of Physics, Concordia University, Montreal, QC H4B 1R6, Canada}}

\author{Maja\ C.\ Cassidy}
\affiliation{School of Physics, The University of New South Wales, Sydney, NSW 2052, Australia}

\author{Susan\ N.\ Coppersmith}
\affiliation{School of Physics, The University of New South Wales, Sydney, NSW 2052, Australia}

\date{\today}

\begin{abstract}
Superconducting nanodevices such as qubits, resonators, and photodetectors, have revolutionized our capabilities for probing and controlling quantum phenomena.
Nonequilibrium quasiparticles, which are broken Cooper pairs that cause decoherence and energy loss, can limit their performance. The number of these quasiparticles is often tens of orders of magnitude greater than would be present in thermal equilibrium.
Background radiation has been shown to excite quasiparticles, but quasiparticles are observed even when the devices are carefully shielded.
Here we show that the high-frequency components of 1/f noise can excite quasiparticle pairs and
that this mechanism is consistent with previously unexplained experimental results.
We also propose new experiments that exploit this
quasiparticle excitation mechanism to non-invasively characterize high-frequency charge noise as well as the locations and nature of the defects producing the noise. The proposed experiments would also investigate how this noise changes as the defects that give rise to it evolve towards thermal equilibrium.
\end{abstract}

\maketitle


\noindent
\textit{Introduction:}
Superconducting circuits underpin quantum computing, sensing, and metrology~\cite{Krantz2019p021318,EsmaeilZadeh2021p190502,Valenti2019p054087}. Yet, their performance is limited by nonequilibrium quasiparticles, which are broken Cooper pairs that cause decoherence and energy loss. 
The persistent presence of quasiparticles at millikelvin temperatures has been a long-standing mystery~\cite{Aumentado2004p066802,Vool2014,Glazman2021p31}.
The leading explanation has been that these quasiparticles arise from background radiation, for instance from infrared radiation~\cite{Corcoles2011p181906,liu2024p017001}, or cosmic rays~\cite{Vepslinen2020p551,Wilen2021p369,Harrington2025p6428}. However, even when extensive efforts to remove these sources of radiation are made, unexplained quasiparticle excitation is observed~\cite{Pan2022p7196}.

In this paper we show that the high-frequency components of 1/f noise can excite quasiparticles in superconducting devices.
This mechanism provides a natural framework for understanding previously mysterious experimental observations.
Such observations include:
\begin{enumerate}
    \item 
    The rate of quasiparticle excitation does not display a significant dependence on the magnitude of the superconducting gap~\cite{Kurter2022p31}.
    \item 
    The rate of quasiparticle excitation can vary greatly between samples that are nominally identical~\cite{Kurter2022p31}.
    \item 
    The presence of a ground plane can greatly reduce the rate at which quasiparticles are excited ~\cite{Mannila2021p145,Higginbotham2022p126}.
    \item 
    When a superconducting device is not subject to an externally applied voltage, the rate of quasiparticle excitation can decrease systematically over a period of months~\cite{Mannila2021p145}.
    \item 
    There are physically reasonable situations in which the rate at which quasiparticles are excited can grow as the size of the device is {\em decreased}~\cite{deRooij2025p024007}.
\end{enumerate}
These observations provide substantial evidence that the mechanism discussed in this paper is observed in experiments.

We also discuss how measurements of quasiparticle excitation rates can be used to gain important information about high-frequency 1/f noise as well as strategies for device designs that can substantially reduce quasiparticle excitation arising from this mechanism.
\newline

{\it Mechanism for quasiparticle excitation by 1/f noise:}
Here we outline the physical picture underlying our calculations.
We assume that 1/f noise arises from charged defects that can fluctuate between two different configurations with similar energies \cite{Anderson1972,Phillips1972,Muller2019}.
These defects are typically far from thermal equilibrium because large voltages are applied to the sample in the course of device operation~\cite{Astafiev2004p267007,Gustafsson2013p245410,Salvino1994,rogge1996p3136}; moreover, the strong interactions between defects cause relaxation to equilibrium to be very slow~\cite{Salvino1994,rogge1996p3136,Carruzzo1994}.
Charge noise has been measured in superconducting devices at frequencies $f$ of order 100~GHz~\cite{Covington2000p5192,Paladino20141p361,Astafiev2004p267007}, where $h f$ is significantly higher than $k_{\rm B} T$, where $T$ is the device temperature.
The noise spectrum can extend out to these high frequencies because the defects that give rise to 1/f noise are often not in thermal equilibrium~\cite{Astafiev2004p267007,Burnett2014p4119}.
Thus it is reasonable to expect that the noise spectrum extends to high enough frequencies~\cite{Quintana2017} that it can excite quasiparticle pairs out of the superconducting condensate.

\begin{figure}
\includegraphics[width=0.5 \textwidth]{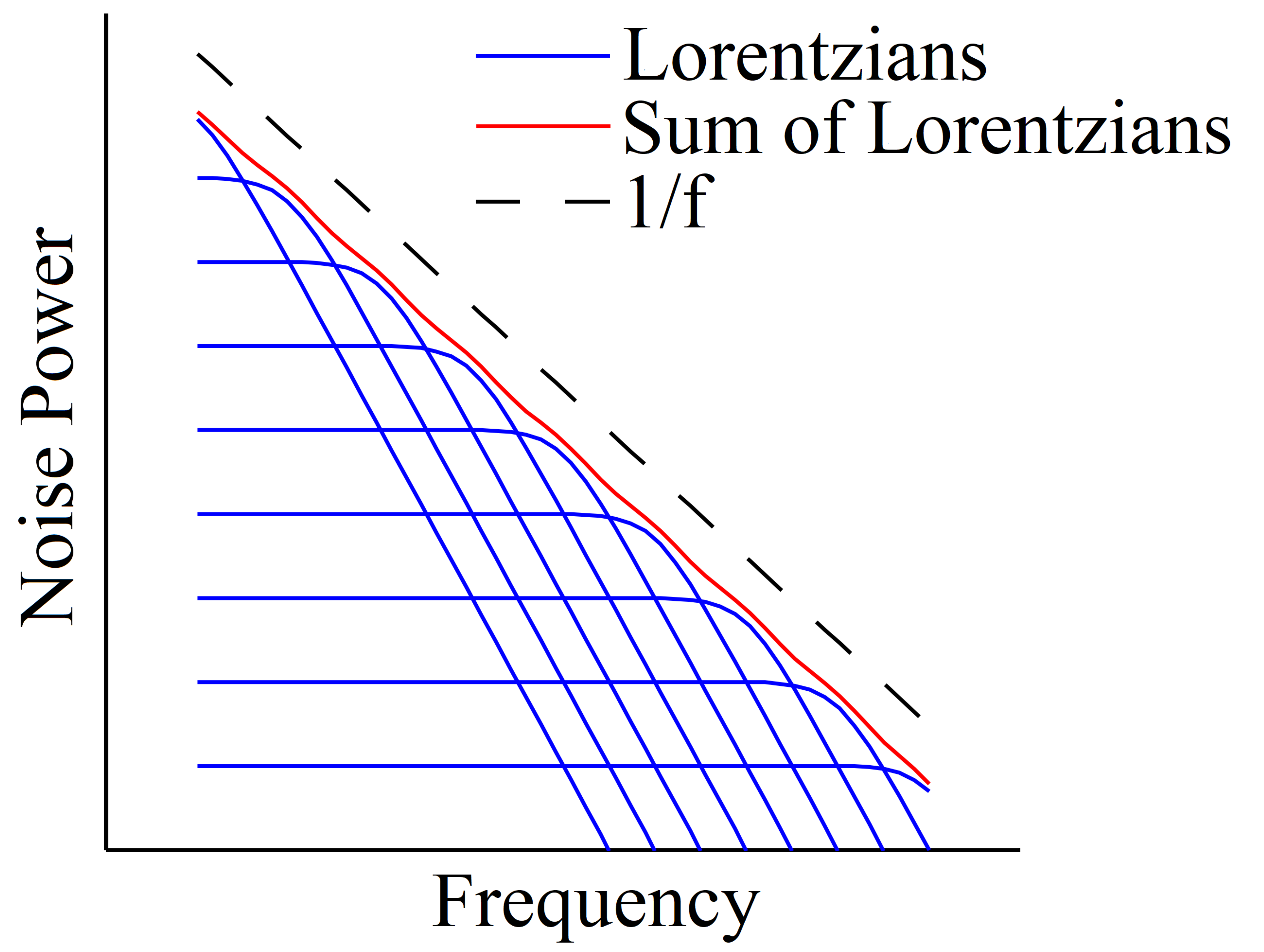}
\caption{
Sum of Lorentzian spectra yields a 1/f noise spectrum.
}
\label{fig:1onfspectrum}
\end{figure}

To calculate the rate of quasiparticle excitation, we consider the bulk excitation of quasiparticle pairs in a translationally-invariant s-wave superconductor with a Hamiltonian, written in $k$-space, of~\cite{marder2010condensed}
\begin{equation}
    H = \sum_k 
    \begin{pmatrix}
        c_{k\uparrow}^\dagger & c_{-k\downarrow}
    \end{pmatrix}
    \begin{pmatrix}
    \epsilon_k & \Delta \\ 
    \Delta^* & -\epsilon_k
    \end{pmatrix}
    \textcolor{black}{
    \begin{pmatrix}
        c_{k\uparrow}  \\c_{-k\downarrow}^\dagger
    \end{pmatrix}},
    \label{eq:Hamiltonian}
\end{equation}
where $c_{k\sigma}^\dagger$ ($c_{k\sigma}$) creates (annihilates) a fermion with spin $\sigma$ and wavevector $k$, $\epsilon_k = \hbar^2 k^2/2m - \mu$, with
$\mu$ the chemical potential,
$\Delta$ is the superconducting pairing amplitude, and the sum is over the first Brillouin zone.
Noise is introduced as a spatially homogeneous but time-varying chemical potential $\mu$~\cite{Mishmash2020p075404}.
The fluctuations of $\mu$ are specified by a noise spectral density $S(\omega)$, which can be written as $S(\omega) = |\mu(\omega)|^2$, where $\mu(\omega)$ is the Fourier transform of $\mu(t)$:
\begin{equation}
    \mu(\omega) = \int_{-\infty}^{\infty} dt ~ e^{-i\omega t} \mu (t)~,
\end{equation}
where $\mu(t)=\mu_0+\delta\mu(t)$. 
The physical origin of the fluctuations $\delta\mu(t)$ is believed to be two-level fluctuators (TLFs)~\cite{Anderson1972,Phillips1972,Grigorij2012,Bilmes2020,Lisenfeld2019p105,Muller2019,Dutta:1981p497,Paladino20141p361,deLeon2021p2823}, which are defects that make stochastic transitions between two states that involve charge motion.
The 1/f spectrum results from summing the Lorentzian spectra of an ensemble of defects with a distribution of transition rates that is uniform on a logarithmic scale, as illustrated in Fig.~\ref{fig:1onfspectrum}.

The chemical potential is spatially homogeneous on the relevant timescales because a localized perturbation from an individual TLF will result in a spatially homogeneous change in the chemical potential of the superconductor on a frequency scale of the plasma frequency of the metal, which is typically orders of magnitude higher than frequencies corresponding to superconducting gaps~\cite{ashcroft1976solid}.

The Hamiltonian Eq.~(\ref{eq:Hamiltonian}) does not couple the different wavevectors $k$.
For each $k$ term in the Hamiltonian matrix there is a ground state $|\Omega_k\rangle$ with energy $-E_k$
and an excited state $|\eta_k \rangle$ with energy $+E_k$, where $E_k=\sqrt{\epsilon_k^2+\Delta^2}$.
The time-dependent perturbation for a given $k$ is $\delta\mu\hat{N}_k$, where $\hat{N}_k =
c_{k\uparrow}^\dagger c_{k\uparrow} + c_{-k\downarrow}^\dagger c_{-k\downarrow}$.
Straightforward calculations similar to those in Sec.~S2A of Ref.~\cite{Alase2025p22394} yield
\begin{equation}
|\langle\eta_k|\hat{N}_k|\Omega_k\rangle|^2 = \left | \frac{\Delta}{E_k} \right |^2~.
\end{equation}
\begin{figure*}
    \centering
\includegraphics[width=1.0\textwidth]{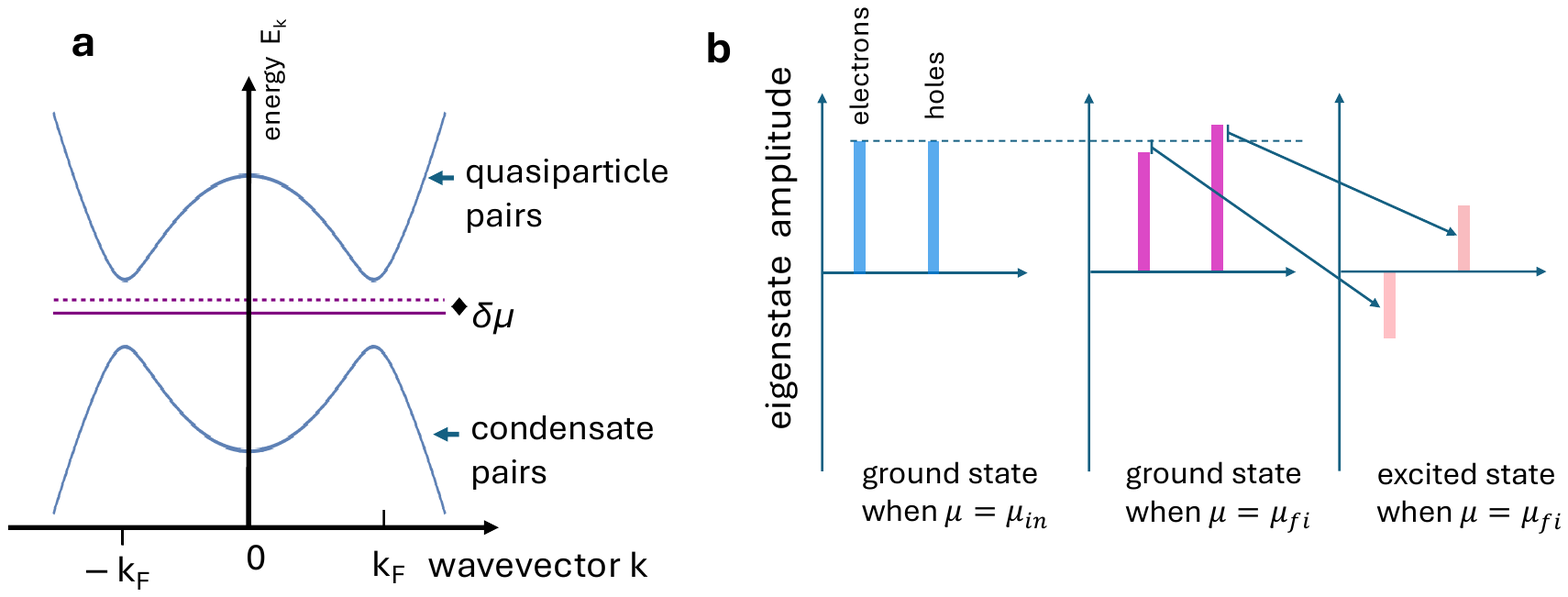}
\vskip -4.5 cm
\caption{
Schematic of energy levels of the superconducting Hamiltonian, Eq.~\ref{eq:Hamiltonian}.
a:  Schematic of energy levels of the superconducting Hamiltonian, Eq.~\ref{eq:Hamiltonian}.
In the absence of noise at zero temperature, the chemical potential $\mu=E_F$, where $E_F$ is the Fermi energy, and the eigenstates of the Hamiltonian at the Fermi level are equal mixtures of electrons and holes.  For each $k$ the ground state is a Cooper pair in the superconducting condensate, while the excited state is a quasiparticle pair.  Changing $\mu$ changes the relative energy cost of holes and electrons. b: Illustration of the change in the wavefunction when the chemical potential is changed at frequencies exceeding the superconducting gap frequency. Quasiparticles must be excited because the relative amounts of electrons and holes in the condensate wavefunction change.
}
\label{fig:schematic_of_calculation}
\end{figure*}
Fig.~\ref{fig:schematic_of_calculation} presents a schematic of the physics underlying the excitation of quasiparticle pairs.
Again, for each $k$, there are two eigenstates; the lower energy one is a condensate pair and the higher energy one is a quasiparticle pair. 
The relative amounts of particles and holes in these eigenstates depend on the value of the chemical potential $\mu$,
so changing $\mu$ quickly causes the condensate pair wavefunction at the old $\mu$ to be a superposition of condensate pairs and quasiparticle pairs at the new $\mu$.
Using methods similar to those in Sec.~S2B of Ref.~\cite{Alase2025p22394}, we
use Fermi's golden rule~\cite{baym2018lectures} to compute the rate $R_{\rm QPP}$ of excitation of quasiparticle pairs at a given k, noting that  $|\delta\mu(\omega)|^2=S(\omega)$ where $S(\omega)$ is the noise power spectrum of the fluctuations in $\mu$~\cite{Mishmash2020p075404}.
We find:
\begin{eqnarray}
    R_{\rm QPP}(k) &=& \frac{2\pi}{\hbar^2} \left | 
    \langle \eta_{k} |\hat{N}_k |
    \Omega_k \rangle
    \right |^2 
    S \left ( \frac{2E_k}{\hbar}\right )
    \\
    &=& \frac{2\pi}{\hbar^2} \left | 
    \frac{\Delta}{E_k}\right |^2
    S \left ( \frac{2E_k}{\hbar}\right )~.
    \label{eq:quasiparticle_rate_one_k}
\end{eqnarray}
For the special case of a 1/f noise spectrum, $S(\omega)=S_0/\omega$, we find
\begin{equation}
    R_{\rm QPP}=\frac{\pi S_0}{\hbar}
    \frac{\left | \Delta \right |^2}{E_k^3}~.
\end{equation}
The total rate of quasiparticle pair excitation is obtained by summing the contributions of all the wavevectors $k$ in the first Brillouin zone.
Appendix~\ref{appendix:quasiparticle_excitation_rate} presents the details of this summation for a thin film with dimensions $L_x$ and $L_y$, which yields the result for the total rate of excitation of quasiparticle pairs, $R_{\rm QPP}$:
\begin{equation}
\label{eq:quasiparticle_rate_2d}
    R_{\rm QPP}= S_0 \frac{L_xL_y E_F}{v_F^2 \hbar^3}~,
\end{equation}
where $E_F$ is the Fermi energy and $v_F$ is the Fermi velocity. Appendix~\ref{appendix:dimensionality} shows that films with thicknesses up to several hundred nanometers are two-dimensional in this context.
%
\\
\\
We now discuss how the quasiparticle excitation rate given in Eq.~(\ref{eq:quasiparticle_rate_2d}) depends on various device parameters.
\\
\\
\textit{Superconducting gap:} First we note that the quasiparticle excitation rate given in Eq.~(\ref{eq:quasiparticle_rate_2d}) does not depend on the magnitude of the superconducting gap.
This lack of dependence on the gap magnitude holds only when the power spectrum of the charge noise is exactly $1/f$.
In Appendix~\ref{appendix:quasiparticle_excitation_rate} we show that when the noise power spectrum decays as $1/f^\alpha$, the rate of quasiparticle excitation is proportional to $\Delta^{1-\alpha}$, where $\alpha$ is the noise exponent.
Thus, it is even possible for the quasiparticle excitation rate to increase as the gap increases, i.e., if the charge noise spectrum decays with frequency more slowly than 1/f.
We note that Ref.~\cite{Astafiev2004p267007} reports a noise spectrum that actually increases at frequencies exceeding $10~\rm GHz$.
If quasiparticles were excited by electromagnetic radiation from cryostat wiring, then one would expect that the quasiparticle excitation rate would consistently decrease as the superconducting gap is increased, because some photons incident on the sample have energies that can excite quasiparticles in low-gap but not high-gap materials.
We note that experimental measurements reported in Fig.~2 of Ref.~\cite{Kurter2022p31} yield a  quasiparticle excitation rate that is somewhat higher for NbN (gap$\sim $2.9 $\mu$eV) than for Al (gap$\sim $0.5 $\mu$eV).
\\
\\
\textit{Variability of quasiparticle excitation rates between nominally identical samples:}
Fig.~2 of Ref.~\cite{Kurter2022p31} also reports quasiparticle excitation rates that vary by over an order of magnitude for devices that are lithographically identical.
Such variability arises naturally when the quasiparticles are excited by 1/f noise, since the 1/f noise depends on the properties of a rather small number of defects near the substrate-metal interface that will vary significantly between samples. 
\\
\\
\textit{Effects of changing device dimension:} 
We now discuss how the dimensions of the device affects the rate of quasiparticle excitation by 1/f noise.
Eq.~(\ref{eq:quasiparticle_rate_2d}) shows that the rate of quasiparticle excitation is proportional to $S_0 A$, where $A$ is the area of the superconductor.
However, we show here that $S_0$ has a nontrivial dependence on A, and that the most reasonable physical models yield a quasiparticle excitation rate that is either constant as a function of device size or else {\em decreases} as A is increased.

To characterize the dependence of $S_0$ on the surface area of a thin film superconductor, we assume that the standard picture of 1/f noise arising from the motion of charges applies~\cite{Dutta:1981p497,deLeon2021p2823}.
The noise arises from the motion of charges in the material, and a charge fluctuation $\delta q$ results in a change in the chemical potential $\delta\mu$ via $\delta\mu =(\delta q)^2/2C$, where $C$ is the capacitance.
Since $C$ is proportional to A, and $S_0$ is proportional to $(\delta\mu)^2$~\cite{Alase2025p22394}, if the magnitude of the charge noise (measured in units of $e/\sqrt{\rm Hz}$, where $e$ is the charge of an electron) is independent of area, then $R_{\rm QPP} \propto A^{-1}$.
However, the area dependence of $S_0$ must also be considered when calculating how the quasiparticle excitation rate depends on area.

We now consider how the fluctuations in charge, $\delta q$, scale with the area of the superconductor.
If we assume that the noise arises from two-level fluctuators (TLFs)~\cite{Dutta:1981p497,deLeon2021p2823} 
and that the correlations between the noise arising from the different fluctuators are negligible, then the spectral density of the ensemble of fluctuators is the sum of the noise power densities of the individual fluctuators in the ensemble.
If the relevant TLFs are distributed uniformly under the superconducting film, then the charge noise power spectrum would be proportional to the area of the superconductor, and the quasiparticle excitation rate would be independent of sample area.
This dependence is consistent with the results reported in Fig.~3c of Ref.~\cite{Pan2022p7196}, where the quasiparticle excitation rate depends much more weakly on sample size when the devices are encapsulated, which markedly lowers the excitation rate due to incident radiation.
However, the active TLFs could be concentrated along the perimeter of the superconductor~\cite{Bilmes2020} (this could plausibly arise because the metal covers the surface of the substrate, which is cleaned before deposition, while the substrate outside the perimeter is exposed to the atmosphere). 
For this case, the charge noise power is proportional to the perimeter of the superconducting area, resulting in a quasiparticle excitation rate that is proportional to $1/\sqrt{A}$, i.e., the quasiparticle excitation rate actually decreases as the sample area is increased.
Interestingly, experiments on superconducting resonators have yielded a quasiparticle excitation rate that increases as the resonator size is decreased~\cite{deRooij2025p024007}.
\\
\\
\textit{Reduction of quasiparticle excitation rate by increased coverage of the substrate with a metallic ground plane:} 
Because charge noise can excite quasiparticle pairs, it is natural to expect that maximizing the coverage of a ground plane, which screens the effects of TLFs that are directly above the ground plane, can reduce the rate of quasiparticle excitation.
It has been noticed previously that increasing coverage by a ground planecan yield samples with anomalously low quasiparticle generation rates~\cite{Mannila2021p145,Higginbotham2022p126}.
\\
\\
\textit{Slow relaxation of quasiparticle excitations in undisturbed devices:}
The TLFs that give rise to 1/f noise interact strongly with each other and exhibit slow relaxations~\cite{Yu1988p231,Carruzzo1994,rogge1996p3136}. 
The rate at which quasiparticles are excited exhibits slow decay over a period of months~\cite{Mannila2021p145}.
 Applying large voltages changes the configuration of the TLFs~\cite{Salvino1994,Carruzzo1994,Pourkabirian2014,Lisenfeld2019p105}, which starts off a new relaxation process at a known time.
This controllability of TLF behavior can be exploited to perform experiments that investigate the correlations between slow relaxations of TLFs and the excitation of quasiparticle pairs.
\\
\\
\textit{Estimate of the rate of quasiparticle excitation by 1/f noise:}
Eq.~(\ref{eq:quasiparticle_rate_2d}) is a formula for the rate of excitation of quasiparticle pairs in terms of the Fermi velocity and Fermi energy of the material along with the power spectrum of the chemical potential fluctuations.
However, the magnitude of the chemical potential fluctuations at frequencies comparable to the superconducting gap frequency is uncertain because the noise is usually measured at low frequencies (often kHz or less) and the extrapolation to high frequencies is uncertain (see, e.g.,~\cite{Christensen2019p140503}).
We make an estimate by taking a charge noise amplitude of $S(\omega)=\alpha/2\omega$ with $\alpha \approx (10^{-3})^2$~\cite{Astafiev2004p267007},
a typical transmon device area  $A=0.26~{\mathrm{mm}}^2$, a capacitance $C=100~{\mathrm{fF}}$~\cite{Martinis2022p26}, a Fermi velocity of 2~x~$10^8$~cm/s and a Fermi energy of 11.6~eV, and we find an excitation rate of about 70 quasiparticle pairs per second. 
This rate is rather reasonable.
\\
\\
\textit{Reducing the rate of quasiparticle excitation caused by 1/f noise:}
The theory described here indicates that quasiparticle excitation may be reduced not just by reducing the charge noise but also by increasing the relevant capacitance, which reduces the magnitude of the chemical potential fluctuations for a given charge fluctuation.
This suggests that connecting the superconductor galvanically to another metal so that the capacitance can be increased without increasing the size of the region in which quasiparticles can be excited could substantially reduce the rate of excitation of quasiparticle pairs.
Of course, when adding more metal to the surface, it is important to control the geometry to avoid creating increased quasiparticle excitation by funneling external radiation into the device~\cite{liu2024p017001}.
\\
\\
\textit{Using quasiparticle excitation to probe high-frequency charge noise:}
Because high-frequency charge noise excites quasiparticle pairs, measurements of quasiparticle generation can provide new information about the nature of $1/f$ noise.
Measuring the rate of quasiparticle excitation while systematically varying the magnitude of the superconducting gap (keeping the surface preparation constant) could yield quantitative information about the noise exponent $\alpha$ since the quasiparticle excitation rate will depend on the magnitude of the gap if $\alpha\neq 1$. The gap may also be varied by changing the temperature or by applying a current, though further work would be needed to disentangle the effects of 1/f noise from the effects of thermal excitation and/or nonequilibrium effects of the current.
\begin{figure}
    \centering
\includegraphics[width=0.4\textwidth]{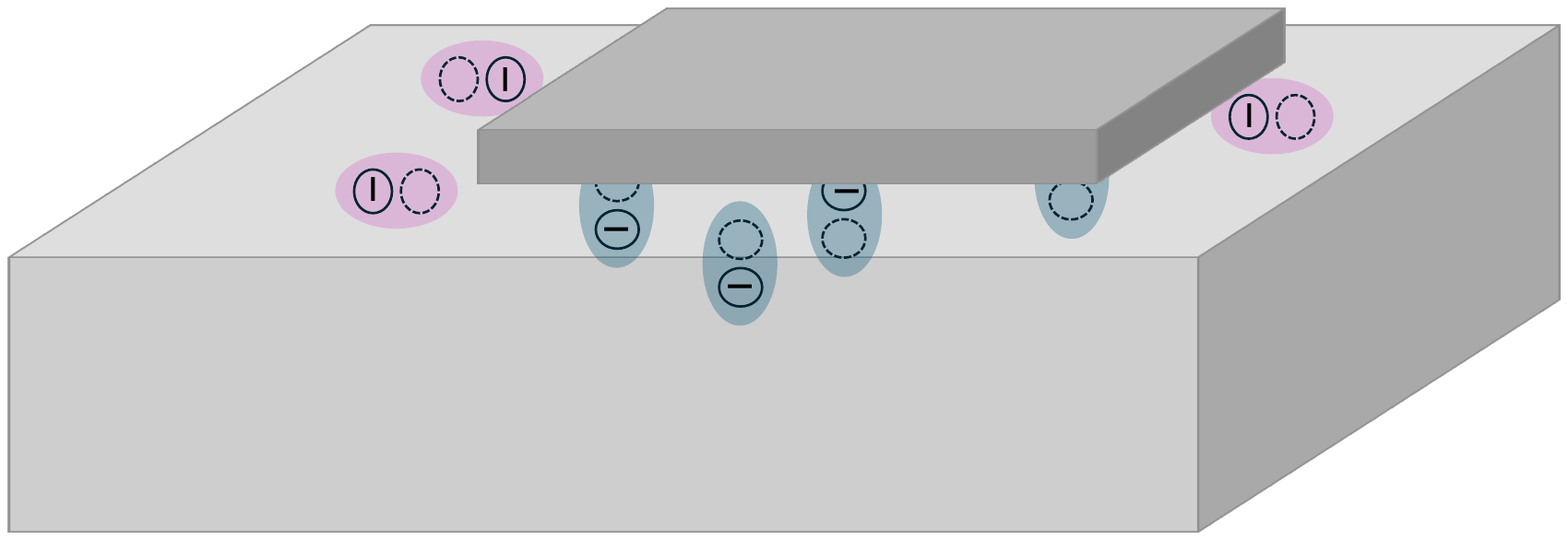}
\vskip -2 cm
\caption{
Schematic illustrating the possibility of obtaining information about the orientation of TLF dipoles by measuring excitation of quasiparticle pairs.
All the dipoles are in the substrate (lighter gray); the blue dipoles are under the superconductor, and are vertically oriented because their fluctuations have a greater effect on the chemical potential of the superconductor.
The fluctuations pink dipoles, which are not directly under the superconductor, have
a larger effect when they are horizontally oriented.
}
\label{fig:dipole_schematic}
\end{figure}
It may also be possible to gain new information about the orientation dependence of the relevant TLFs.
The reason for this is illustrated in Fig.~\ref{fig:dipole_schematic}; under the superconductor, the TLFs that generate the largest chemical potential fluctuations are oriented vertically, while in the exposed areas, the TLFs that change $\mu$ the most are oriented horizontally (unless they are under a ground plane, in which case the image charges in the ground plane will screen out the fields of the horizontal dipoles).
%
\newline

\textit{Conclusions:} We have shown that quasiparticles can be excited in superconducting devices by 1/f noise.
The mechanism that produces quasiparticles yields a rate of excitation that depends on sample size, superconducting gap magnitude, and time in ways that differ from previously proposed excitation mechanisms. In addition,
there are reports in the literature of experimental measurements consistent with our 1/f-noise mechanism.
Our results imply that new insights into 1/f noise can be obtained by careful measurements of quasiparticle excitations in superconducting devices.
\newline

\textit{Acknowledgments:}
We acknowledge useful correspondence with Xuedong Hu and John Martinis.
Work at UNSW was supported by the U.S.\ Army Research Office under Award No. W911NF-23-1-0110, and by Google Asia Pacific Pte.\ Ltd.  M.C.G acknowledges additional support from the Sydney Quantum Academy.
A.A.\ acknowledges support by the Australian Research Council Centre of Excellence for Engineered Quantum Systems (Grant No.\ CE170100009). M.C.C.\ acknowledges support from a UNSW Scientia Fellowship and an Australian Research Council Discovery Early Career Research Fellowship (Grant No.\ DE240100590). C.C.Y.\ acknowledges support from the U.S. Army Research Office in collaboration with the Laboratory for Physical Sciences (Grant No.\ W911NF24C0001).
The views, conclusions, and recommendations contained in this document are those of the authors
and are not necessarily endorsed by nor should they be
interpreted as representing the official policies, either expressed or implied, of the Army Research Office or the
U.S. Government. 
The U.S.\ Government is authorized to
reproduce and distribute reprints for U.S.\ Government
purposes notwithstanding any copyright notation herein.
\newline

\appendix

\section{Derivation of quasiparticle excitation rate using Fermi's golden rule}
\label{appendix:quasiparticle_excitation_rate}
We begin with the standard BCS Hamiltonian for a translationally invariant s-wave superconductor in the main text, which
we present again here for convenience:
\begin{equation}
    H = \sum_k 
    \begin{pmatrix}
        c_{k\uparrow}^\dagger & c_{-k\downarrow}
    \end{pmatrix}
    \begin{pmatrix}
    \epsilon_k & \Delta \\ 
    \Delta & -\epsilon_k
    \end{pmatrix}
    \begin{pmatrix}
        c_{k\uparrow} &c_{-k\downarrow}^\dagger
    \end{pmatrix},
\end{equation}
where $c_{k\sigma}^\dagger$ ($c_{k\sigma}$) creates (annihilates) a fermion with spin $\sigma$ and wavevector $k$, $\epsilon_k = \hbar^2 k^2/2m - \mu$, with
$\mu$ the chemical potential,
$\Delta$ is the superconducting pairing amplitude, and the sum is over the first Brillouin zone.
Noise is incorporated as a time-varying chemical potential~\cite{Mishmash2020p075404}, $\mu(t) = \mu_0 + \delta\mu(t)$~\cite{Mishmash2020p075404}.
We calculate the rate of quasiparticle pair excitation using Fermi's golden rule.
We assume the power spectral density of these chemical potential fluctuations is $S(\omega) = S_0/\omega$.

As discussed in the main text, straightforward calculation yields an expression for $R_{\rm QPP}(k)$, the probability of exciting a quasiparticle pair at one value of $k$ (Eq.~\ref{eq:quasiparticle_rate_one_k}): 
\begin{eqnarray}
    R_{\rm QPP}(k) &=& \frac{2\pi}{\hbar^2} \left | 
    \langle \eta_{k} |\hat{N}_k |
    \Omega_k \rangle
    \right |^2 
    S \left ( \frac{2E_k}{\hbar}\right )
    \\
    &=& \frac{2\pi}{\hbar^2} \left | 
    \frac{\Delta}{E_k}\right |^2
    S \left ( \frac{2E_k}{\hbar}\right )~.
    \label{eq:quasiparticle_rate_one_k_in_appendix}
\end{eqnarray}

To obtain the total rate of quasiparticle pair excitation, $R_{\rm QPP}$, we sum over the $k$'s in the first Brillouin zone.
We show in Appendix~\ref{appendix:dimensionality} that typical superconducting devices including transmon qubits and superconducting resonators are well-described as two-dimensional systems, so here we perform the summation for the two-dimensional case.

We convert the sum over $k$-vectors to an integral in the usual way, $\sum_k \rightarrow L_x L_y (2\pi)^2\int d^2 k$, where $L_x$ and $L_y$ are the sample dimensions along the in-plane directions $x$ and $y$.
We have
\begin{equation}
    R_{\rm QPP} = \frac{L_x L_y}{(2 \pi)^2}(2 \pi) \int_0^\infty dk ~k
    \left [ \frac{\pi {S_0} \Delta^2}{\hbar E_k^3} \right ]~.
\end{equation}

We assume that the superconducting gap is much smaller than temperature and that the temperature is much smaller than the Fermi energy of the metal and write $E_k = \hbar v_F(k-k_F)$ and $\mu \rightarrow E_F$.
Making the variable substitution
$z = \hbar v_F(k-\mu/\hbar v_F)/\Delta$ and extending the lower limit of the integration range over z from $-\mu/\Delta$ to $-\infty$ yields the result
\begin{equation}
    R_{\rm QPP} = S_0 \frac{L_xL_y E_F}{\hbar^3 v_F^2}~.
    \label{eq:R_QPP}
\end{equation}

The method presented here is easily generalized to situations where the spectral density of the noise is not 1/f, so that $S(\omega)=S_0/\omega^\alpha$.
Implementing the steps presented above yields the result
\begin{equation}
    R_{\rm QPP} = L_x L_y \sqrt{\pi}\frac{\Gamma(\frac{1+\alpha}{2})}{\Gamma(1+\frac{\alpha}{2})}
    S_0 \frac{E_F}{2\hbar^3 v_F^2}
    \Delta^{1-\alpha}~,
\end{equation}
where $\Gamma(x)$ is the Gamma function.

\section{Thickness at which a superconducting film is two-dimensional for calculations of quasiparticle excitation rate by 1/f noise}
\label{appendix:dimensionality}
Eq.~\ref{eq:quasiparticle_rate_one_k} in the main text shows the rate at which quasiparticle pairs are excited for a single $k$-vector;
this rate is significant only when the energy of the excitation is not too much greater than the superconducting gap.
To determine whether a dimension $L$ of a sample is small enough that only the lowest $k$ vector contributes significantly to the sum over $k$-vectors, we compare the difference in energies of the lowest and second-lowest $k$-vectors to the size of the superconducting gap.
The difference in the energies of the two lowest $k$-vectors is $2\pi \hbar v_F/L$, so only one $k$-vector contributes significantly if $\hbar v_F/L \ll \Delta$, 
or $L \gg 2\pi \hbar v_F/\Delta$.
For a superconducting gap of 1~meV and a Fermi velocity of $10^8$~cm/s, this condition yields a length of about 4~$\mu$m, which is substantially thicker than most superconducting films used in modern devices.



\bibliography{main}

\end{document}